\documentclass[twocolumn,aps,floats,prl]{revtex4}
\usepackage[T1]{fontenc}
\usepackage[latin9]{inputenc}
\usepackage{bm}
\usepackage{amsmath}
\usepackage{amssymb}
\usepackage{wasysym}
\usepackage{graphicx}
\usepackage{amsfonts}
\usepackage{esint}
\usepackage{bm}
\usepackage{dsfont}

\newcommand{\beq}{\begin{equation}} 
\newcommand{\eeq}{\end{equation}} 
\newcommand{\beqa}{\begin{eqnarray}} 
\newcommand{\eeqa}{\end{eqnarray}}

\begin{document}
\bibliographystyle{naturemag}

\title{Collapse of  Landau levels in Weyl semimetals}
\author{Vicente Arjona$^1$, 
Eduardo V. Castro$^{2,3}$, and Mar\'ia A. H. Vozmediano $^1$ }
\affiliation{$^1$ Instituto de Ciencia de Materiales de Madrid, and CSIC, Cantoblanco, 28049 Madrid, Spain}
\affiliation{$^2$ CeFEMA, Instituto Superior T\'{e}cnico, Universidade de 
Lisboa, Av. Rovisco Pais, 1049-001 Lisboa, Portugal}
\affiliation{$^3$Beijing Computational Science Research Center, Beijing 100084, China }

\begin{abstract}
It is known that in two dimensional relativistic Dirac systems, the Landau levels can collapse in the presence of a critical in-plane electric field. We extend this mechanism to the three dimensional Weyl semimetals and analyze the physical consequences for the cases of both, real and pseudo Landau levels arising form strain--induced elastic magnetic fields.

\end{abstract}
\maketitle

\section{Introduction}
Graphene and Weyl semimetals (WSM) are examples of Dirac matter in two and three dimensions respectively. In both materials, the Fermi surface consists on a series of pairs of points of definite chirality, located at different positions in the Brillouin zone. The relativistic nature of these systems has important experimental consequences. One of the best explored in the literature is the behavior in magnetic fields. The Landau level (LL) spectrum differs from that of the standard electron systems and the LL structure was the compelling evidence  of having Dirac electrons in graphene~\cite{Netal05,Zetal05}. The characteristic zeroth Landau level LL plays an important role in the discussion of the chiral anomaly in WSM~\cite{NN83} in the 3D systems and the magneto-resistance has become the standard test of the anomaly~\cite{Bur15,JXH16,Lietal15,HZetal15,XKetal15,ZXetal16}. 

More recently it has been recognized that, as happens in graphene, elastic lattice deformations couple to the low energy electronic excitations of WSM in the form of elastic axial gauge fields~\cite{CFLV15}. The original derivation was followed by a number of works  extracting the consequences of this new vector coupling~\cite{PCF16,GVetal16,LPF17,VGetal16,Getal16}. Strain engineering can  give rise to pseudo LL in these 3D materials, nice examples are provided in refs.~\cite{PCF16,LPF17} and a general analysis of strain induced LL in any dimension has been presented in~\cite{RAV16}. 

In this work we analyze the spectrum of WSM in the presence of perpendicular  electric and magnetic fields. The motivation lies on the recognition that, for Dirac materials, the presence of a critical electric field perpendicular to the magnetic field induces a collapse of the Landau levels. This fact was explicitly derived in two spacial dimensions for the case of graphene in~\cite{LSB07,PC07}, and has also been recognized in a different context in~\cite{YYY16,TCG16} when analyzing  tilted WSM. Our main contribution concerns the strained WSM. The critical observation is that, in addition to the elastic gauge fields, strain gives rise in general to a deformation potential that acts as a pseudo--electric field perpendicular to the pseudo--magnetic field leading to the collapse of the pseudo Landau levels (PLL). This phenomena has been discussed for graphene in~\cite{CCV16}. To fix the notation we will first derive the condition for the collapse of LL in the case of real electromagnetic fields in Sec.~\ref{sec_emc} and then examine the situation of the electronic oscillations induced by elastic deformation in Sec.~\ref{sec_PLL}. We end in Sec.~\ref{sec_discuss} with some further considerations on the effects of strain in WSM. 

\section{Collapse of the Landau levels in Weyl semimetals. Electromagnetic fields} 
\label{sec_emc}
\begin{figure}
\begin{centering}
\includegraphics[width=0.8\columnwidth]{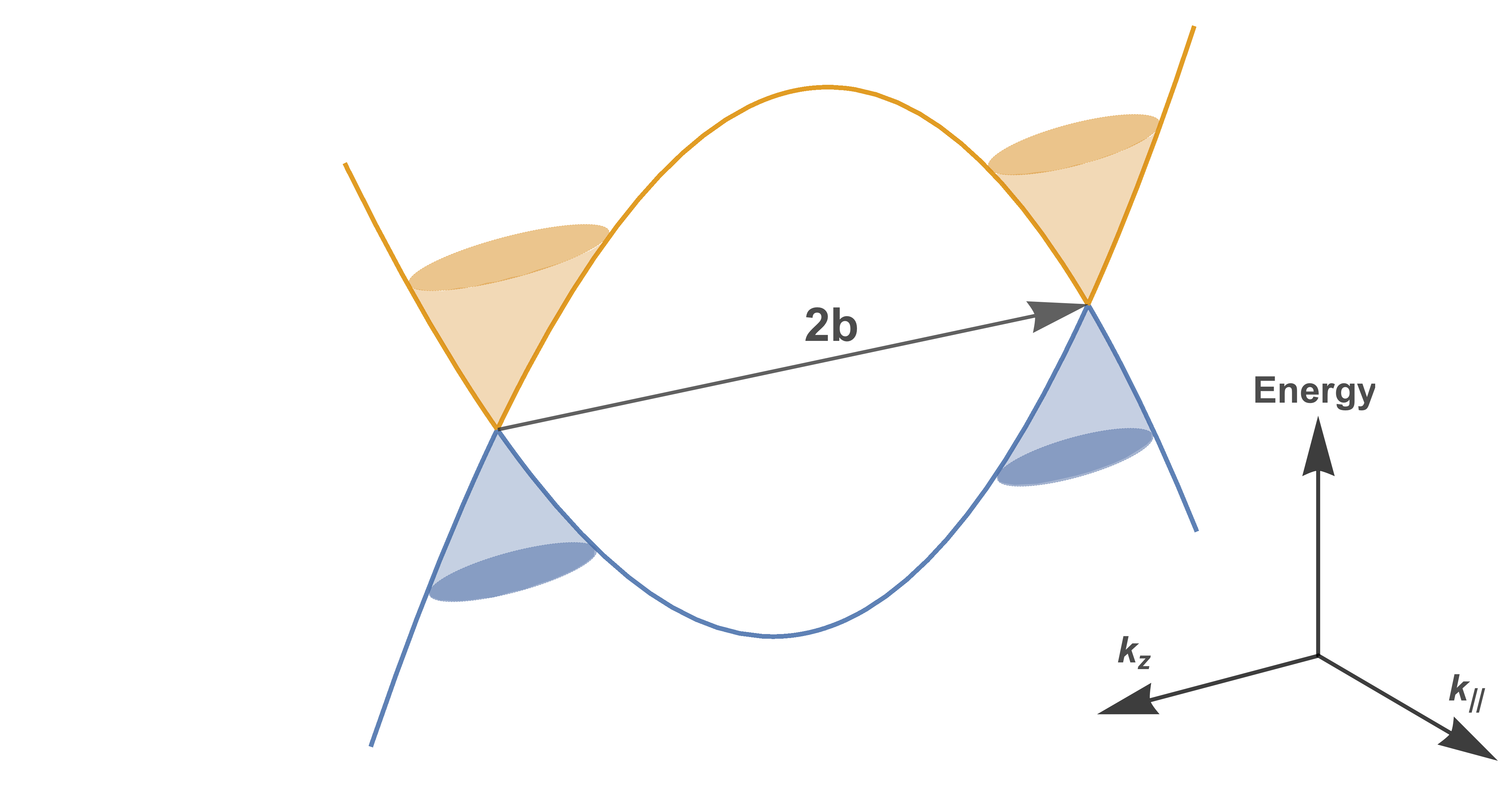}
\par
\end{centering}
\caption{\label{Fig_WSM} Schematic representation of the band structure of eq. \eqref{HWSM} with the Weyl nodes separation chosen along the $k_z$ axis. $k_\parallel$ stands for $(k_x, k_y)$.}
\end{figure}
Dirac fermions in three dimensions are described by a four component spinor wave function obeying the Dirac equation $(\gamma^\mu\partial_\mu-m)\Psi=0$. In the massless case and choosing the Weyl representation for the Dirac matrices, the equation splits into two equations with two-component wave functions representing Weyl fermions of oposite chirality. In the material realization of the Weyl physics, the two chiralities (band crossings) are separated in $k$ space by a vector $b^\mu (\mu=0,1,2,3)$.
The minimal low energy effective model describing a WSM with only two nodes, is
\beq
{\cal L}=\bar\Psi\gamma^\mu(\partial_\mu+i\gamma^5 b_\mu)\Psi,
\label{HWSM}
\eeq 
where $\gamma^\mu$ are standard Dirac matrices.  The vector $b_\mu$  breaks Lorentz symmetry. Its time component $b_0$ is related to the separation in energy of the Weyl nodes and breaks, in addition, inversion symmetry. The spacial component $b_i$ represents the separation in $k$ space and breaks time reversal symmetry $\mathcal{T}$. Fig.~\ref{Fig_WSM} shows a schematic representation of the band structure of eq. ~\eqref{HWSM} with the vector $\vec{b}$ pointing in the $z$ direction. 

Inversion-broken WSM have been found experimentally~\cite{Lvetal15,XuLiuetal15,ZXetal16}; $\mathcal{T}$ broken have been predicted in magnetic compounds~\cite{CSetal16} and experimental evidences are advocated in~\cite{BEetal15}. 

Signatures of the chiral anomaly are found in the behavior of the magnetorresistance in parallel E and B fields. The case of perpendicular E and B fields has been less studied although a  very interesting discussion of the LL spectrum  in the context of tilted Weyl semimetals is done in  refs.~\cite{YYY16,TCG16}. The LL collapse is already implicit  there but their aim is different.  We will present here the derivation  of the LL collapse in perpendicular electric and magnetic field to fix the notation and to pave the way for the discussion of strain in the next section.

\begin{figure}
\begin{centering}
\includegraphics[width=0.9\columnwidth]{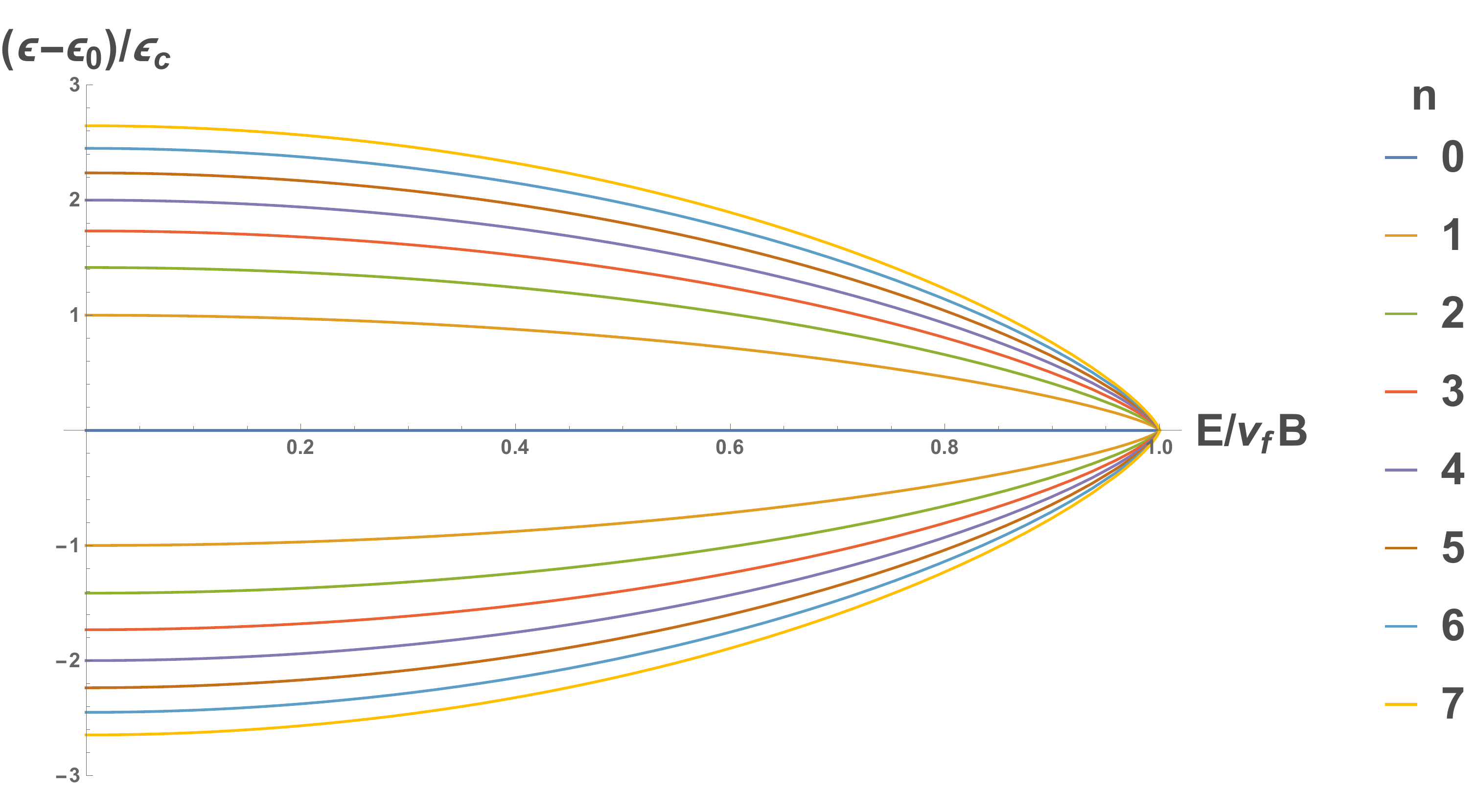}
\par
\end{centering}
\caption{\label{Fig_cyclo} Cyclotron frequency as a function of the electric field for $k_z$=0. 
}
\end{figure}

Around a single Weyl node, the WSM is a relativistic system with the velocity of light $c$ replaced by the Fermi velocity $v_F$. As we know from special relativity~\cite{LL71,J98}, a boost in the direction perpendicular to E and B with the appropriate velocity leads to a reference frame  where the electric field $E^\prime$ vanishes. The spectrum of a WSM in perpendicular E and B is obtained by solving the problem in the primed reference frame with magnetic field $B^\prime$ and boosting back to the original reference frame. This was done in the 2D case in ref. \cite{LSB07}. The 3D derivation is as follows: 

We choose the Landau gauge  $\mathbf{A}=(-By,0,0)$ and the scalar potential $\phi=-Ey$. This represents a constant magnetic  and electric fields $\bf{B,E}$ pointing in the $z$ and $y$ directions respectively. Since $k_z$ is a good quantum number, the system can be considered as a collection of 2D Dirac layers in a perpendicular electromagnetic field. The energy spectrum of such configuration is well known; electron motion is arranged into Landau levels: 
\beq 
\epsilon_n=\pm \sqrt{\Omega_c^2  n +v_f^2 k_z^2 },
\label{3DLL}
\eeq 
where the cyclotron frequency in units $\hbar=1$ is $\Omega_c=\sqrt{2}v_F/l_B$, with $l_B = \sqrt{1 /(eB)}$ the magnetic length. 
Under a boost in the $x$ direction with velocity $v$, the chosen electromagnetic field transforms as:
\beq
E'_y=\gamma (E_y-vB_z) \quad,\quad B'_z=\gamma(B_z-\frac{v}{v_F^2}E_y) ,
\eeq
where $\gamma=1/\sqrt{1-\beta^2}$, $\beta=v/v_ F$. When the velocity coincides with the drift velocity, $v_d=E/B$, the primed reference frame experiences only a magnetic field of magnitude $B'_z=\sqrt{1-\beta^2}B$ and the spectrum is given by eq.~\eqref{3DLL} with the primed magnetic field and with $\beta=E/v_FB$. Since the energy is the zeroth component of the energy-momentum quadrivector, the inverse boost transformation 
\beq 
\epsilon_n= \gamma(1-\beta^2) \epsilon'_n+v_f \beta k_x
\eeq 
gives the spectrum in the original frame. The final expression is
\beq 
\epsilon_n=\pm  \sqrt{\Omega_c^2  n (1-\beta^2)^{3/2}+v_f^2  k_z^2 (1-\beta^2)}+v_f\beta k_x .
\label{E-B}
\eeq 
For comparison, the spectrum of a non-relativistic electron system in the same field configuration is
\beq
\epsilon_n=\left(n+\frac{1}{2}\right)\omega_c+\frac{k_z^2}{2m}- k_y\frac{E}{B}-\frac{m}{2}\left(\frac{E}{B}\right)^2,
\label{egas}
\eeq
with the cyclotron frequency given by the standard expression $\omega_c=eB/mc$. As it happens in the 2D case~\cite{LSB07,PC07}, apart from a rigid shift of the levels (which is the only effect of the electric field in the case of the non-relativistic electron system), there is a non trivial dependence of the cyclotron frequency with the electric field.
The evolution of the cyclotron frequency with the applied electric field  for $k_z=0$ is shown in Fig.~\ref{Fig_cyclo}.
As it can be seen, the LL collapse for the critical value $E=v_F B$.
The LL collapse described here is different from the overlap of LL occurring in a non relativistic electron gas~\cite{KBK14}. It is a property of the Dirac matter that will take place not only in WSM but also in Dirac semimetals. These materials have the two chiral nodes located at the same point in the Brillouin zone and are, at present, more accesible experimentally. Examples are $Na_3Bi$ or $Cd_3As_2$ \cite{Liuetal14,JZetal14,XKetal15}. Three dimensional Dirac LL have already been observed in these compounds so it would be interesting to check the LL collapse in these materials. Although indications of LL collapse have been obtained experimentally \cite{SD09,GRetal11} we do not know of similar attempts in 3D samples.
%

\section{Strain in Weyl semimetals: Collapse of the pseudo Landau levels}
\label{sec_PLL}
\begin{figure}
\includegraphics[width=1.0\columnwidth]{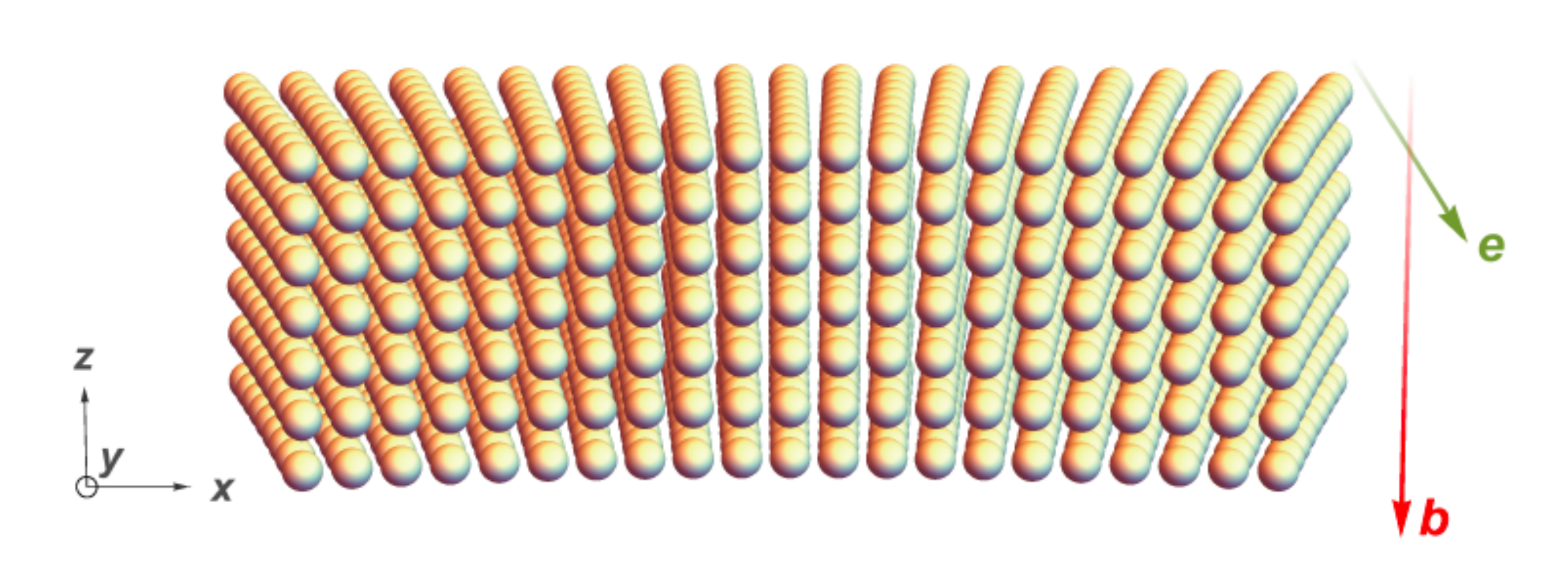}
\caption{\label{deform} Sketch of a sheet of the proposed rectangular lattice WSM. The applied strain configuration bends the original system into a circular arc  in the $x$-$y$ plane. This configuration is able to produce both pseudo--magnetic and pseudo--electric fields, the PLL spectrum collapsing when the critical value is reached. } 
\end{figure}
The coupling of electronic excitations to lattice deformations in graphene in the form of gauge fields~\cite{KM97,SA02b,VKG10}, the prediction of pseudo Landau levels (PLL) for some strain configurations~\cite{GKG10,GGKN10}, and their experimental observation with Scanning Tuneling Spectroscopy~\cite{LBetal10} have been one of the most exotic and fruitful developments in the physics of graphene. The concept of elastic gauge fields and the capability to manipulate the electronic excitations by means of strain (straintronics) has been later extended to other more general 2D Dirac materials~\cite{Amorim16}. Massless Dirac nodes must come in pairs of opposite chirality~\cite{NN81}. The general understanding is that elastic gauge fields will arise when these Dirac nodes  are located at non equivalent points in the Brillouin zone~\cite{CFLV16}. 
Weyl semimetals share most of the relevant properties of graphene concerning the coupling of elastic and electronic degrees of freedom. 
The fact that elastic gauge fields would also be present in WSM was deduced in a recent publication~\cite{CFLV15}. In this section we will explore the collapse of the PLL in certain strain configurations due to the strain induced deformation potential following the lines described for graphene in~\cite{CCV16}.

Elastic lattice deformations are parametrized by the strain tensor $u_{ij}=1/2(\partial_i u_j+\partial_j u_i)$, a function of the displacement vector $u_i$. A symmetry approach allows a straightforward construction of an elastic gauge field as $A_i^{el}=u_{ij} b_j$ ($i,j=x,y,z$). The coupling parameters must be fixed either by a microscopic model (tight binding, {\it ab initio}) or determined by experiments. In the original reference~\cite{CFLV15}, these were extracted from a tight-binding model.

Another important electron-phonon coupling in elasticity theory is the deformation potential (shifts in the energy bands resulting from deformations of the crystal lattice). It is proportional to the trace of the strain tensor and is of the same order as the elastic gauge field in a derivative expansion. Considering the simplest case of two Weyl nodes separated in momentum space, the low energy Hamiltonian around one of the Weyl nodes coupled to the lattice deformation is
\beq
{\cal H}=v_F\sigma^i(p_i+c A_i)+g\Phi \mathds{1}, 
\label{Hstr}
\eeq
where $\Phi(x)\sim \sum_ju_{jj}(x)$ and g is the coupling constant associated to the deformation potential. The dimensionless Gr\"{u}neisen parameter c is typically of order one in most materials \cite{HM02,IL09}. In what follows we will put  $c=1$  and remove it from the discussion. The deformation potential couples as a scalar electromagnetic  potential which can give rise to a (pseudo) electric field $E_i=-\partial_i\Phi$. Next we will demonstrate the collapse of the PLL in a very simple configuration and later we will discuss more realistic examples.

Assume that we have a WSM with two Weyl nodes separated a distance $b$ in the $k_x$ direction.
A strain configuration such that $u_{xx}=-By/b, u_{yy}=u_{zz}=0$, will give rise to an elastic vector potential $A_i=u_{ij}b_j$ such that $A_x=-By$, $A_y = A_z=0$. This describes a uniform magnetic field of magnitude $B$ in the $z$ direction. Simultaneously, the scalar potential $\Phi=-gyB/b$, induces a pseudo--electric field of magnitude $E=gB/b$ pointing in the $y$ direction. We are then in the situation described in the previous section. The collapse of the PLL occurs when $E\geq v_FB$, i. e. $g B/b\geq v_F B$. The condition for the PLL collapse translates into a constraint on the values of the  coupling constant associated to the deformation potential: $g\geq v_F b$. Although there are not yet measurements of the electron--phonon couplings in WSM, using reasonable numbers we see that the condition will be easily attainable in the samples. The separation between the nodes is estimated to be $b\sim 0.08 \text{\AA}^{-1}$ in $TaAs$, with the Fermi velocity $v_F\sim 1.3\times 10^5 m/s$~\cite{HXetal17}. These values give a lower bound for the elastic coupling constant of $g\geq 0.07 eV$ meaning that the samples will typically be affected by the effect described. Notice that the estimated value  in graphene  is of the order of 4 eV taking screening into account. Thin films of WSM which are the ones more suitable for straintronics are not expected to be so deformable but the required value is fifty times smaller and screening is reduced in WSM due to the relativistic spectrum.


Apart from the academic example presented to expose the phenomenon, more realistic strain configurations will also be affected by it and this effect will have to be taken into account for a correct interpretation of the experiments when these are available. Since the deformation potential is the trace of the strain tensor, shear strain conserving the volume of the sample will not generate a pseudo--electric field. This is the case for example of the torsional strain discussed in~\cite{PCF16} or the tetra-axial strain in the diamond lattice of ref.~\cite{RAV16}. Other strain configurations easier to implement experimentally will be fully affected by the effect. 
The best experimentally accessible devices will be obtained by bending thin films of WSM, a generalization of the strain configuration first suggested in ref.~\cite{GGKN10} for graphene sheets. Consider a rectangular lattice model with the Weyl nodes separated by a distance $b$ in the $k_x$ direction.  A generic deformation 
\begin{eqnarray}
u_x&=&u_0(2 x y + Cx)  \nonumber \\
u_y&=&u_0[-x^2-Dy(y+C)] \nonumber \\
u_z&=&0 ,
\label{paco2}
\end{eqnarray}
where $u_0$ and $D$ are constants that depend on the material; $u_0$ defines the maximum stress and  $D$ is a relation between Lam\'e coefficients. $C$ parametrizes a family of deformations encoding the same pseudomagnetic field. The strain configuration \eqref{paco2} gives rise to the elastic gauge potential $A_x=u_{xx}b_x= 
u_0(2y+C)b$. The constant pseudo--magnetic field will be $B_z=-2 u_0 b$ . In addition to the pseudo--magnetic field, this strain configuration has a deformation potential 
$\Phi(y)=u_0(1-D)(2y+C)$ that will generate a constant pseudo--electric field $E_y=-2 u_0 (1-D) g$ perpendicular to the magnetic field, as shown in Fig.~\ref{deform}. As in the above example, the collapse of the PLL spectrum is translated into a restriction for the values of the elastic coupling constants, $g\geq v_f b /(1-D)$. For a thin film of $Cd_3As_2$ as the one suggested in ref.~\cite{LPF17}, a coupling constant $g \geq 0.16 eV$,  would be needed to collapse the predicted oscillations. 
 

\section{Discussion}
\label{sec_discuss}
This work can be extended to situations involving more general -- and more realistic -- materials. In particular, most of the actual WSM are inversion broken, meaning that the inequivalent Weyl nodes are separated in energy by a zeroth component of the $b$ field in eq.~\eqref{HWSM}. In a recent publication~\cite{CKLV16} it was shown that, in these materials, a time component of the elastic gauge field will develop under strain. This gives rise to a pseudo--electric gauge field that, in contrast to the one associated to the deformation potential, will be axial, i. e. will couple with opposite sign to the two Weyl nodes of opposite chirality. The addition of the two terms can lead to interesting situations where the chiral imbalance is maximized by making the total scalar potential  zero in one of the nodes. 

The  Dirac semimetals have been the subject of intense experimental research and the measures of the magneto-resistance have been used as an experimental evidence of the chiral magnetic effect~\cite{FKW08,LKetal16}. Even though these materials will not support elastic gauge fields, strain will still induce a deformation potential that will also affect the spectrum of the system in real magnetic fields.

\vspace{0.3cm}
\begin{acknowledgments}
We thank A. Cortijo, A. G. Grushin, F. de Juan and Y. Ferreiros for useful conversations. Special thanks go to J. Silva-Guill\'en and P. San-Jose for help with the figures.
This work  has been partially supported by Spanish MECD
grant FIS2014-57432-P,  the Comunidad de Madrid
MAD2D-CM Program (S2013/MIT-3007), and  FCT-Portugal through Grant No.~UID/CTM/04540/2013.

\end{acknowledgments}

\bibliography{WSM}
\end{document}